# Emergence of Quantum Confinement in Topological Kagome Superconductor $CsV_3Sb_5$ family

Yongqing Cai[1*], Yuan Wang[1*], Zhanyang Hao[1*], Yixuan Liu[1*], Xiao-Ming Ma[1], Zecheng Shen[1], Zhicheng Jiang[2], Yichen Yang[2], Wanling Liu[2], Qi Jiang[2], Zhengtai Liu[2], Mao Ye[2], Dawei Shen[2], Zhe Sun[3], Jiabin Chen[4], Le Wang[1], Cai Liu[1], Junhao Lin[1], Jianfeng Wang[5,4#], Bing Huang[4], Jia-Wei Mei[1#] and Chaoyu Chen[1#]

[1] Shenzhen Institute for Quantum Science and Engineering (SIQSE) and Department of Physics, Southern University of Science and Technology (SUSTech), Shenzhen 518055, China.

[2] State Key Laboratory of Functional Materials for Informatics and Center for Excellence in Superconducting Electronics, Shanghai Institute of Microsystem and Information Technology, Chinese Academy of Sciences, Shanghai 200050, China.

[3] National Synchrotron Radiation Laboratory, University of Science and Technology of China, Hefei, Anhui 230029, China

[4] Beijing Computational Science Research Center, Beijing 100193, China

[5] School of Physics, Beihang University, Beijing 100191, China

[*] These authors contributed equally to this work.

[#]Correspondence should be addressed to J.W. (wangjf06@buaa.edu.cn), J.M. (meijw@sustech.edu.cn) and C.C. (chency@sustech.edu.cn)

**Abstract**

**Quantum confinement is a restriction on the motion of electrons in a material to specific region, resulting in discrete energy levels rather than continuous energy bands. In certain materials quantum confinement could dramatically reshape the electronic structure and properties of the surface with respect to the bulk. Here, in the recently discovered kagome superconductor $AV_3Sb_5$ ($A$=K, Rb, Cs) family of materials, we unveil the dominant role of quantum confinement in determining their surface electronic structure. Combining angle-resolved photoemission spectroscopy (ARPES) measurement and density-functional theory simulation, we report the observations of two-dimensional quantum well states due to the confinement of bulk electron pocket and Dirac cone to the nearly isolated surface layer. The theoretical calculations on the slab model also suggest that the ARPES observed spectra are almost entirely contributed by the top two layers. Our results not only explain the disagreement of band structures between the recent experiments and calculations, but also suggest an equally important role played by quantum confinement, together with strong correlation and band topology, in shaping the electronic properties of this family of materials.**

**Introduction:**

The observations of superconductivity and chiral charge density wave (CDW) orders in $A$V$_3$Sb$_5$ ($A$ = K, Rb, Cs) family of materials [1-4] have provoked intensive research interest in condensed matter physics. The critical temperature of superconductivity was found as: $T_C = 0.93$ K for KV$_3$Sb$_5$ [5], $T_C = 0.92$ K for RbV$_3$Sb$_5$ [6], and $T_C = 2.5$ K for CsV$_3$Sb$_5$ [1], respectively. Meanwhile, a chiral CDW order was observed at $T \sim 80 - 103$ K [1-3,5-8], whose relationship with superconductivity is still under debate [3,9-15]. Recent experimental and theoretical works have suggested the unconventional nature of both orders [3,7,14-18]. While the nature and relationship between these two orders are under active investigations [1-18], the nontrival band topology inherited from their two-dimensional (2D) kagome lattice [2,5,6] is certainly playing a fundamental role.

Partially inspired by their $2 \times 2 \times 2$ CDW order, the experimental studies on the electronic structures of this family have been mainly concentrated on the CDW gap and van Hove singularities (VHSs) [1,7,13,19-27]. General agreement has been reached on the formation of momentum-dependent CDW gaps [19,20,26,27] as directly measured by angle-resolved photoemission spectraoscopy (ARPES). Concerning the VHSs, while several ARPES works have reported the existence of multiple VHSs around the $\overline{M}$ point with pure or mixed sublattice nature [23,25,26], such observations are only partially in line with the density-functional theory (DFT) calculations [1,28,29]. To be specific, DFT has predicted two VHSs in a narrow energy region ($< 200\ meV$) below the Fermi level at the $M$ point of bulk but no VHS in the same energy window at the $L$ point because of a clear $k_z$ dispersion along $M - L$. The recent ARPES works [23,26] have reported the observations of these two VHSs through a conjucture guided by DFT, yet their $k_z$ information and the band structure at the $L$ point are missing. In fact, most of the ARPES spectra at $\overline{M}$ [1,7,13,19-27] are close to the energy bands at the $L$ point of bulk rather than the $M$ point predicted by DFT calculations. Such disagreement between experiment and theory not only calls for a systematic ARPES experiment to examine the $k_z$ dispersion, but also an alternative picture to correctly understand the observed electronic structures.

In this letter, we report an universal quantum confinement phenomemon at the surface of kagome superconductor $A$V$_3$Sb$_5$ family of materials, which can fully account for the observed spectral features, including the lack of $k_z$ dispersion as discussed above, and the emergence of quantum well states which we discovered for the first time in this family. Through systematic photon energy dependent ARPES measurements, we observe a weak but periodic $k_z$ dispersion along the $\Gamma - A$ direction but absence of $k_z$ dispersion along the $M - L$ and $K - H$ directions. This causes the results that the measured ARPES spectra along both $K - M - K$ and $H - L - H$ look almost identical to the DFT calculated bands along $H - L - H$. Furthermore, a clear band splitting is observed for the electron pocket at $\overline{\Gamma}$, leading to a subband with pure 2D nature. Both of these features suggest the formation of quantum well states due to the confinement of bands on the surface. Simulated by a slab model with 6 layers of CsV$_3$Sb$_5$, all the ARPES spectral features can be captured by our DFT calculations considering the surface relaxation effect. It is found that the interlayer spacing on the surface is increased, especially for that between the topmost and the second kagome layers, which results in a nearly isolated surface layer and the 2D band structure. The ARPES spectra are thus mainly composed of the quantum well states residing on the top two kagome layers. Our results unveil the dominant role of quantum confinement, together with strong correlation and band topology, in shaping the electronic properties of $A$V$_3$Sb$_5$ family of materials.

**Results:**

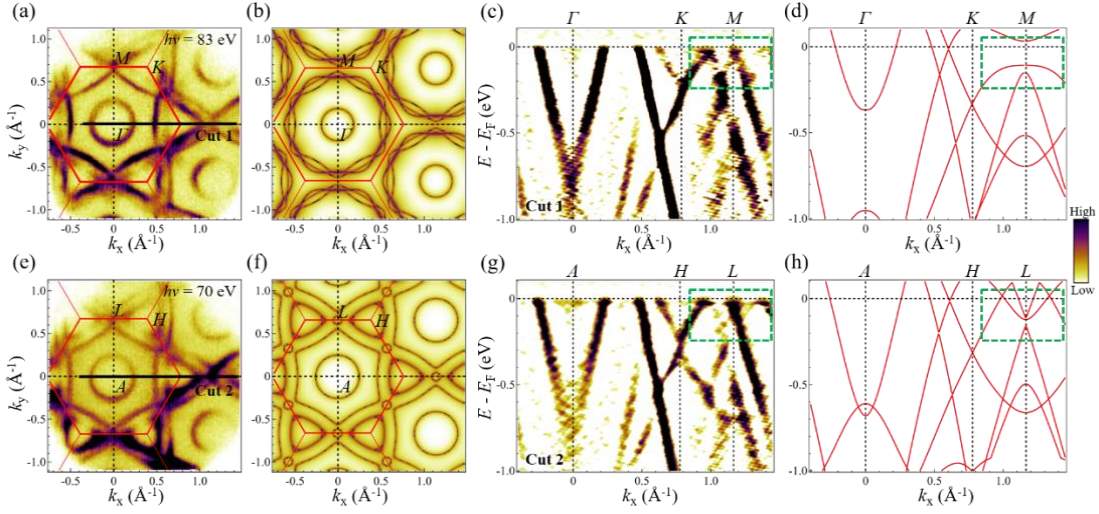

**Fig. 1. Comparison between ARPES measurements and DFT calculated band structures of CsV$_3$Sb$_5$.** (a) Fermi surface mapping of CsV$_3$Sb$_5$ measured with photon energy $h\nu = 83\ eV$, which corresponds to $k_z = 7 * 2\pi/c$ plane. (b) DFT calculated constant energy contour (CEC) at $k_z = 0$ plane. (c) Second derivative plots of the band structure along $\Gamma - K - M$ as indicated in (a) by the black solid line. (d) DFT calculated band structure along the $\Gamma - K - M$ direction. (e-h) Same as (a-d) but measured at $k_z = 6.5 * 2\pi/c$ (photon energy $h\nu = 70\ eV$) and calculated at $k_z = \pi$ plane, respectively. The hexagonal Brillouin zones are marked by the red solid lines.

We first present the comparison between the ARPES spectra and DFT calculated band structures of bulk, and show their main disagreements as encountered in most recent works. Taking CsV$_3$Sb$_5$ as an example, photon energy dependent ARPES measurements have been performed. According to the periodic spectral intensity variation of energy distribution curves (EDCs) at the $\bar{\Gamma}$ point, the high-symmetry points $\Gamma$ and $A$ of bulk are determined, which correspond to photon energies of $83\ eV$ (Fig. 1(a)) and $70\ eV$ (Fig. 1(e)), respectively. Experimental and calculated electronic structures are compared at the $\Gamma - M - K$ plane (Figs. 1(a-d)) and $A - L - H$ plane (Figs. 1(e-h)). For the DFT calculated Fermi surfaces, clear differences can be seen at the $\Gamma - M - K$ and $A - L - H$ planes. For example, while the $A - L - H$ Fermi surface (Fig. 1(f)) shows closed loops surrounding the bulk $H$ point, they expand in the $\Gamma - M - K$ plane, merging with each other and forming a closed loop surrounding the bulk $\Gamma$ point (Fig. 1(b)). The tiny loops surrounding the bulk $L$ point (Fig. 1(f)) are absent at the bulk $M$ point (Fig. 1(b)). In sharp contrast, despite some intensity variation, the ARPES measured Fermi surfaces are almost identical to each other at both the $\Gamma - M - K$ (Fig. 1(a)) and $A - L - H$ (Fig. 1(e)) planes, both of which are close to the DFT calculated one at the $A - L - H$ plane (Fig. 1(f)).

Such differences between ARPES measurements and DFT calculations can be also reflected by the band structures along high-symmetry lines. According to DFT results, there is a small gap between conduction and valence bands at the bulk $M$ point (dashed box in Fig. 1(d)), while two conduction and one valence bands cross with each other near the Fermi level around the bulk $L$ point (dashed box in Fig. 1(h)). In particular, the outer conduction band centered at $L$ crosses one branch of the upper Dirac cone centered at $H$, forming a new Dirac node almost at the Fermi level. This DFT predicted $k_z$ dependence along $M-L$ is missing in the ARPES results, as one can see that both ARPES spectra along $\Gamma-K-M$ (Fig. 1(c)) and $A-H-L$ (Fig. 1(g)) show very simiar features. The Dirac nodes at the Fermi level as discussed above can be clearly observed in both ARPES spectra along $K-M$ and $H-L$. In summary, the DFT calculations on the bulk $CsV_3Sb_5$ predict a clear $k_z$ dispersion along $M-L$, but ARPES results around the $\overline{M}$ point are $k_z$ independent, with spectral features almost identical to the DFT calculated results around the bulk $L$ point.

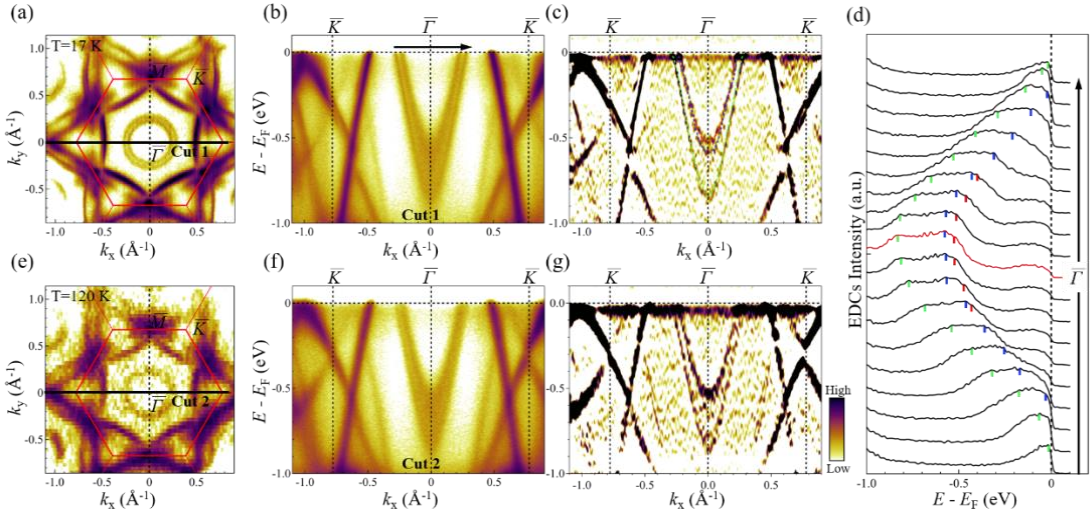

**Fig. 2. Emergence of quantum well states from the electron pocket at $\overline{\Gamma}$.** (a) Fermi surface mapping of $CsV_3Sb_5$ measured with $77\,eV$ photons at 17 K. (b) Band structure measured along the $\overline{\Gamma}-\overline{K}$ direction marked in (a) by the black solid line. (c) The second derivative image of (b). (d) EDCs stack plot of (b) with the black arrows in (b) and (d) marking the momentum region and direction. Red, blue and green tips mark the peak position of the three bands around $\overline{\Gamma}$ that are also plotted in (c) by the red, blue and green dashed lines. (e-g) Same as (a-c) but measured at 120 K, above the CDW transition temperature.

We then move to the electron pocket around the $\overline{\Gamma}$ point. Before discussing its $k_z$ dispersion, we will study its detailed band structure based on the high-quality ARPES data we could achieve in this family of materials. Figure 2 presents the ARPES results measured by a general photon energy ($77\,eV$). Different from the single electron pocket predicted by DFT (Figs. 1(b, d, f, h)), the ARPES results show two circles around $\overline{\Gamma}$ at the Fermi surface (Fig. 2(a)), two main electron-like bands crossing the Fermi level along the high-symmetry cut $\overline{\Gamma}-\overline{K}$ (Figs. 2(b, c)), and multiple EDC peaks in the corresponding energy-momentum window. These features of multiple bands seem to be the quantized subbands of the bulk electron pocket, which are reminiscent of the quantum well states observed in topological insulators [30-33]. Interestingly, simiar band splitting phenomena are also observed by ARPES in another material $KV_3Sb_5$ (see Supplementary Fig. S2), indicating the universality of quantum well states in this family.

Other trivial scenarios may also account for band duplicates. The first one is surface inhomogeneity, which could lead to multiple sets of bands from different domains. This is not the present case as the subbands observed here are all symmetric with respect to $\overline{\Gamma}$, and not all the main bands have replicas. The second possibility is superlattice bands induced by CDW, which can be also ruled out as these subbands are still present above the CDW transition temperature (Figs. 2(e-g)). The third one is caused by $k_z$ projection, in which the minimum and maximun edge of band continuum along $k_z$ contribute two band-like features in ARPES spectra [34]. Such features have indeed been captured in recent ARPES measurement from this family of materials [19,21,22,25]. However, the energy difference between the splitted bands we observed (~250 $meV$) is much smaller than the $k_z$ band width of this electron pocket (~400 $meV$). Furthermore, the number of splitted bands we find here is more than just two, as proved by weak features in Fig. 2(c) indicated with blue dashed lines and peaks in Fig. 2(d) by blue tips. Both features suggest that $k_z$ projection is unlikely the reason of our case. Other possibilities such as polaronic effect [35,36] can also be excluded since these splitted bands cross the Fermi level. The above analyses strongly point to quantum well states as the origin of these splitted bands.

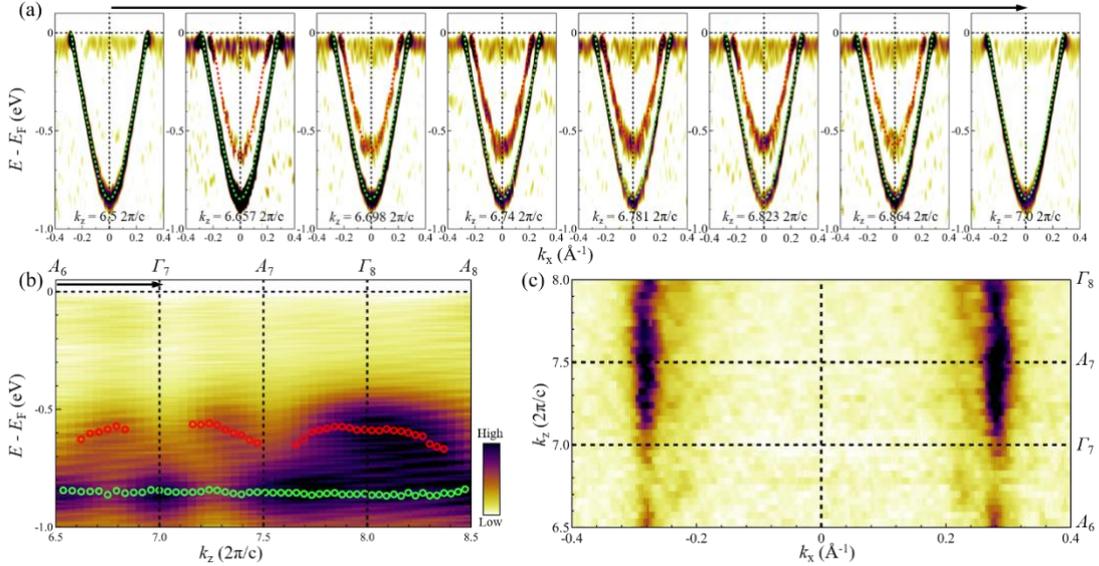

**Fig. 3. $K_z$ dispersion of the quantum well states.** (a) Band structure around Brillouin zone center $\overline{\Gamma}$ under a series of photon energies. Main subbands are labelled by the red and green dashed lines, in the same way as in Fig. 2(c). (b) Photon energy depedent spectra for photoemission intensity at the Brillouin zone center. Open circles represent the binding energy evolution of the corresponding subbands in (a). (c) ARPES spectral intensity map around the Fermi level in $k_x - k_z$ plane. The two strong linear features correspond to the Fermi momenta of the outer subband.

The definitive evidence of quantum well states come from the (quasi-) 2D nature of the splitted subbands. In Fig. 3(a), we show the photon energy dependent spectra, corresponding to 8 consecutive $k_z$ values between $k_z = 6.5 * 2\pi/c$ and $k_z = 7.0 * 2\pi/c$. The most striking feature is the outmost electron-like band as highlighted by the green dashed lines. At the first glance, this electron band shows no detectable change with $k_z$. By detailed analysis, the band bottom and Fermi vectors all present flat dispersion with $k_z$, as demonstrated in Figs. 3(b) and (c), respectively. This clearly proves the outmost subband as a surface quantum well state. Concerning the inner electron subband, the same analyses presented in Figs. 3(a) and (b) unveil its $k_z$ dependence with a band width $\sim 100\ meV$. However, this $k_z$ dispersion is much weaker than the bulk band ($\sim 400\ meV$) calculated by DFT. It suggests that besides the bulk nature, the inner subband also shares a similar origin of quantum confinement, but its confinement is weaker than that of the outmost band. All these measurements indicate that a well potential may be formed on the surface.

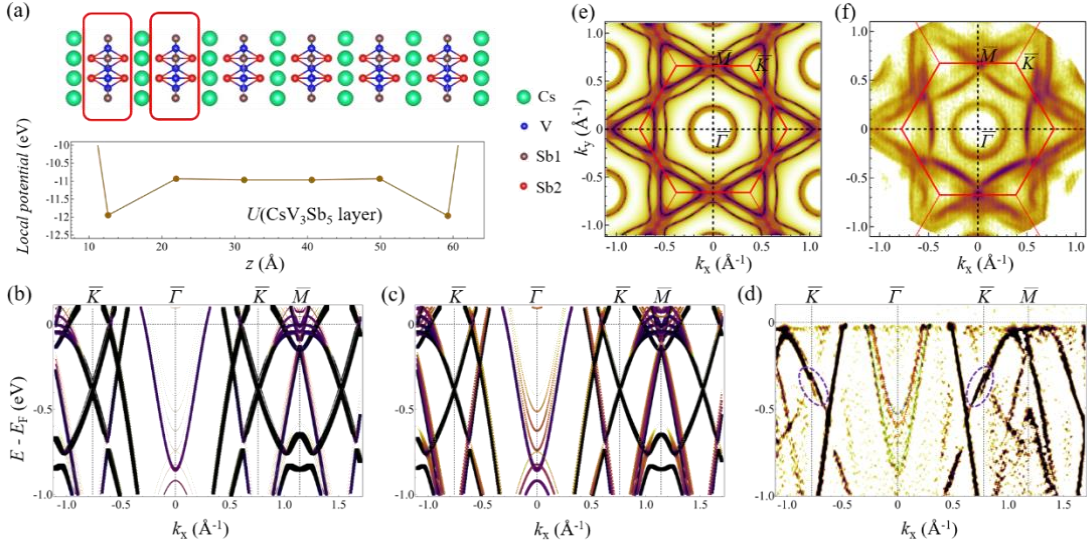

**Fig. 4. Surface relaxation effect on the formation of quantum well states.** (a) Upper panel: schematic of slab structure consisting of six layers of $CsV_3Sb_5$. Lower panel: average local potential on each $CsV_3Sb_5$ unit layer along the $z$ direction of slab after surface relaxation. (b, c, d) Comparison of calculated and measured band structures along the $\overline{K} - \overline{\Gamma} - \overline{K}$ direction. In (b) and (c), the simulated band structures are projected to the topmost layer and the top two layers, respectively. In (d), the ARPES data was measured using $77\ eV$ photon energy, same as that in Fig. 2. (e, f) Comparison of calculated and measured Fermi surface.

To demonstrate the origin of this quantum confinement, we build a symmetric slab model consisting of six $Cs-Sb_2-V/Sb_1-Sb_2$ layers of $CsV_3Sb_5$ with Cs atomic layers at both terminations (upper panel of Fig. 4(a)). The (0001) surface of $CsV_3Sb_5$ belongs to a polar surface. In general, a stable polar surface of solid is often accompanied by large structural relaxation or surface reconstruction [37]. Our DFT calculations reveal that the structural relaxations mainly occur at the surface layers of $CsV_3Sb_5$ (see Supplementary Fig. S4). The topmost $Cs-Sb_2-V/Sb_1-Sb_2$ unit layer is found far away from the units below, and atomic-layer spacing inside this topmost unit is reduced, which leads to the surface unit layer of $CsV_3Sb_5$ relatively free-standing. In contrast, the relaxation effect of inner units is negligible and they almost keep the bulk structure nature. In the lower panel of Fig. 4(a), we plot the average local potential on each $CsV_3Sb_5$ unit layer along the $z$ direction of slab. Remarkably, a potential drop ($\sim 1\ eV$) occurs at the surface due to the relaxation, which causes the formation of a quantum confinement near the surface. In Figs. 4(b) and (c), we present the DFT calculated band structures of slab with projections onto the topmost unit and the top two units, respectively. By comparing with ARPES results (Fig. 4(d)), the following relevant messages can be obtained. First, the pure surface quantum well state (the outmost electron band) is almost entirely contributed by the topmost $CsV_3Sb_5$ unit layer (Fig. 4(b)), confirming its origin of quantum confinement. Second, just using the DFT calculated bands from the contributions of the top two unit layers is enough to simulate the ARPES spectra satisfactorily (Fig. 4(c)), in agreement with the ARPES probe depth ($1\sim2\ nm$) within the photon energy range reported. This indicates that the ARPES observed spectra are almost entirely of surface origin, providing consistent explanation for the lack of $k_z$ dispersion at the $\overline{K}$ and $\overline{M}$ points, as well as the weak $k_z$ dependence of the inner electron band at $\overline{\Gamma}$. Finally, the 2D origin of the ARPES observed spectra can be further strengthened by the Dirac cone splitting at the $\overline{K}$ point. As highlighted by dashed ellipses in Fig. 4(d), the Dirac cone which forms a Dirac nodal line along $k_z$ direction is also confined to the topmost layer, composing a new type of topological surface state Dirac cone with quantum well state nature. Employing the slab model of CsV$_3$Sb$_5$, the calculated Fermi surface with projection to the top two unit layers is shown in Fig. 4(e), which is in good agreement with the experimental measurements (Fig. 4(f)) using a general photon energy. Importantly, this Fermi surface is nearly $k_z$ independent due to the (quasi-) 2D nature of bands, significantly different from the cases shown in Figs. 1(b) and (f).

**Discussion:**

We have conducted a deep investigation on the origin of ARPES observed spectra in $A$V$_3$Sb$_5$ ($A$=K, Rb, Cs) family of materials. Likely universal quantum well states, originating from quantum confinement of the corresponding bulk states, are observed in this material family. Using CsV$_3$Sb$_5$ as an example, the formation of quantum well states can be well simulated in a slab model. The quantum confinement arises from the surface relaxation effect of the polar surface, which results in a relatively isolated surface unit layer of CsV$_3$Sb$_5$. The pure 2D quantum well states observed by ARPES are almost entirely residing on the topmost layer. Comparison between experimental and simulated data suggests that the ARPES spectral intensity is dominated by contribution from only the top two layers. Our results not only explain the previous discrepancy between ARPES observations and DFT calculations, but also highlight the important role of quantum confinement in shaping the surface electronic structures of this material family. Besides the strong correlation and band topology, the quantum confinement should be also considered in any attempts to understand the novel surface-related properties of these materials, such as the chiral CDW, the mysterious behaviour of VHSs, and the intrinsic channel of giant anomalous Hall conductivity.

**ACKNOWLEDGEMENTS**

This work is supported by National Natural Science Foundation of China (NSFC) (Grants Nos. 12074163 and 12004030), the Shenzhen High-level Special Fund (Grants No. G02206304 and G02206404), the Guangdong Innovative and Entrepreneurial Research Team Program (Grants Nos. 2017ZT07C062 and 2019ZT08C044), Shenzhen Science and Technology Program (Grant No. KQTD20190929173815000), the University Innovative Team in Guangdong Province (No. 2020KCXTD001), Shenzhen Key Laboratory of Advanced Quantum Functional Materials and Devices (No. ZDSYS20190902092905285), Guangdong Basic and Applied Basic Research Foundation (No. 2020B1515120100), and China Postdoctoral Science Foundation (2020M682780). The authors acknowledge the assistance of SUSTech Core Research Facilities. The calculations were performed at Tianhe2-JK at Beijng Computational Science Research Center.

**Materials and Methods**

**Sample growth and characterization**

KV$_3$Sb$_5$ and CsV$_3$Sb$_5$ single crystals were grown by the self-flux method. High purity K and Cs (clump), V (powder) and Sb (ball) in the ratio of 2:1:6 were placed in an alumina crucible. The crucible was then double sealed into an evacuated quartz tube under a pressure of 10$^{-6}$ Torr to prevent the assembly from reacting with air during the reaction. The quartz tube containing the alumina crucible was placed in the furnace and the temperature was increased to 500 °C in 5 hours. The mixtures was kept in 500 °C for 10 hours. Then the quartz tube was heated at 1050 °C for 12 hours, after this, the quartz tube was cooled to 650 °C very slowly. After the heat treatment, the quartz tube was removed and centrifuged. A large number of millimeter-sized hexagonal KV$_3$Sb$_5$ and CsV$_3$Sb$_5$ single crystals were obtained.

**Transport, heat capacity and magnetic measurements**

Resistivity and heat capacity were measured by Physical Property Measurement System (PPMS, Quantum Design). To characterize the resistivity, a standard four - probe method was employed with a probe current of 0.5mA in the *ab* plane of the crystals. And to measure the heat capacity, several flat crystals of approximate 1.5mm size were selected, and Apezion N-grease was used to ensure the connection with heat capacity stage. Magnetization measurements were performed by Magnetic Property Measurement System (MPMS3, Quantum Design). Quartz paddle was used, and the magnetic field was perpendicular to *c* - axis to minimize the diamagnetic contribution.

**ARPES measurement**

ARPES measurements were performed at the BL03U beamline of the Shanghai Synchrotron Radiation Facility (SSRF) and beamline 13U of the National Synchrotron Radiation Laboratory (NSRL). The energy resolution was set at 15 meV for Fermi surface mapping and 7.5 meV for band structure measurements. The angular resolution was set at 0.1°. Most samples were cleaved *in situ* under ultra-high vacuum conditions with pressure better than $5 \times 10^{-11}$ mbar and temperatures below 20 K, while some samples were cleaved above 120 K.

**First-principles calculations**

The first-principles calculations are performed using the Vienna ab initio simulation package [38] within the projector augmented wave method [39] and the generalized gradient approximation of the Perdew-Burke-Ernzerhof [40] exchange-correlation functional. The plane-wave basis with an energy cutoff of 400 eV is adopted. To simulate the surface structures, the slab model is built, and the thickness of vacuum is taken to be 16 Å. $\Gamma$-centered 12×12×6 *k*-point meshes and 12×12×1 *k*-point meshes are adopted for bulk and slab structures respectively. Employing the experimental lattice constants of *a*=*b*=5.495 Å (and *c*=9.308 Å for bulk), the crystal and slab structures of CsV$_3$Sb$_5$ is relaxed with van der Waals correction [41] until the residual forces on each atom is less than 0.005 eV/Å.

[1] B. R. Ortiz, S. M. L. Teicher, Y. Hu, J. L. Zuo, P. M. Sarte, E. C. Schueller, A. M. M. Abeykoon, M. J. Krogstad, S. Rosenkranz, R. Osborn, R. Seshadri, L. Balents, J. He & S. D. Wilson. $CsV_3Sb_5$: A $Z_2$ Topological Kagome Metal with a Superconducting Ground State. *Phys Rev Lett* **125**, 247002, doi:10.1103/PhysRevLett.125.247002 (2020).

[2] Brenden R. Ortiz, Lídia C. Gomes, Jennifer R. Morey, Michal Winiarski, Mitchell Bordelon, John S. Mangum, Iain W. H. Oswald, Jose A. Rodriguez-Rivera, James R. Neilson, Stephen D. Wilson, Elif Ertekin, Tyrel M. McQueen & Eric S. Toberer. New kagome prototype materials: discovery of $KV_3Sb_5$, $RbV_3Sb_5$, and $CsV_3Sb_5$. *Physical Review Materials* **3**, doi:10.1103/PhysRevMaterials.3.094407 (2019).

[3] Y. X. Jiang, J. X. Yin, M. M. Denner, N. Shumiya, B. R. Ortiz, G. Xu, Z. Guguchia, J. He, M. S. Hossain, X. Liu, J. Ruff, L. Kautzsch, S. S. Zhang, G. Chang, I. Belopolski, Q. Zhang, T. A. Cochran, D. Multer, M. Litskevich, Z. J. Cheng, X. P. Yang, Z. Wang, R. Thomale, T. Neupert, S. D. Wilson & M. Z. Hasan. Unconventional chiral charge order in kagome superconductor $KV_3Sb_5$. *Nat Mater*, doi:10.1038/s41563-021-01034-y (2021).

[4] Shuo-Ying Yang, Yaojia Wang, Brenden R Ortiz, Defa Liu, Jacob Gayles, Elena Derunova, Rafael Gonzalez-Hernandez, Libor Šmejkal, Yulin Chen & Stuart SP Parkin. Giant, unconventional anomalous Hall effect in the metallic frustrated magnet candidate, $KV_3Sb_5$. *Science Advances* **6**, eabb6003 (2020).

[5] Brenden R. Ortiz, Paul M. Sarte, Eric M. Kenney, Michael J. Graf, Samuel M. L. Teicher, Ram Seshadri & Stephen D. Wilson. Superconductivity in the $Z_2$ kagome metal $KV_3Sb_5$. *Physical Review Materials* **5**, doi:10.1103/PhysRevMaterials.5.034801 (2021).

[6] Qiangwei Yin, Zhijun Tu, Chunsheng Gong, Yang Fu, Shaohua Yan & Hechang Lei. Superconductivity and Normal-State Properties of Kagome Metal $RbV_3Sb_5$ Single Crystals. *Chinese Physics Letters* **38**, doi:10.1088/0256-307x/38/3/037403 (2021).

[7] H. X. Li et al. Observation of Unconventional Charge Density Wave without Acoustic Phonon Anomaly in Kagome Superconductors $AV_3Sb_5$ (A=Rb,Cs). *arXiv: 2103.09769* (2021).

[8] Zuowei Liang. Three-dimensional charge density wave and robust zero-bias conductance peak inside the superconducting vortex core of a kagome superconductor $CsV_3Sb_5$. *arXiv: 2103.04760* (2021).

[9] B. Q. Song et al. Competing superconductivity and charge-density wave in Kagome metal $CsV_3Sb_5$: evidence from their evolutions with sample thickness. *arXiv: 2105.09248* (2021).

[10] Feng Du et al. Interplay between charge order and superconductivity in the kagome metal $KV_3Sb_5$. *arXiv: 2102.10959* (2021).

[11] Yanpeng Song et al. Competition of superconductivity and charge density wave in selective oxidized $CsV_3Sb_5$ thin flakes. *arXiv: 2105.09898* (2021).

[12] Han-Shu Xu et al. Multiband superconductivity with sign-preserving order parameter in kagome superconductor $CsV_3Sb_5$. *arXiv: 2104.08810.* (2021).

[13] Rui Lou et al. Charge-Density-Wave-Induced Peak-Dip-Hump Structure and Flat Band in the Kagome Superconductor $CsV_3Sb_5$. *arXiv: 2106.06497* (2021).

[14] C. C. Zhao et al. Nodal superconductivity and superconducting domes in the topological Kagome metal $CsV_3Sb_5$. *arXiv: 2102.08356* (2021).

[15] K. Y Chen, N. N Wang, Q. W Yin, Y. H Gu, K. Jiang, Z. J Tu, C. S Gong, Y. Uwatoko, J. P Sun, H. C Lei, J. P Hu & J. G. Cheng. Double Superconducting Dome and Triple

Enhancement of Tc in the Kagome Superconductor CsV3Sb5 under High Pressure. *Physical Review Letters* **126**, doi:10.1103/PhysRevLett.126.247001 (2021).

[16] F. H. Yu et al. Concurrence of anomalous Hall effect and charge density wave in a superconducting topological kagome metal. *arXiv: 2102.10987* (2021).

[17] Zhiwei Wang et al. Anomalous transport and chiral charge order in kagome superconductor CsV3Sb5. *arXiv: 2105.04542* (2021).

[18] Hui Chen et al. Roton pair density wave and unconventional strong-coupling superconductivity in a topological kagome metal. *arXiv: 2103.09188* (2021).

[19] Kosuke Nakayama et al. Multiple Energy Scales and Anisotropic Energy Gap in the Charge-Density-Wave Phase of Kagome Superconductor CsV3Sb5. *ArXiv: 2104.08042* (2021).

[20] Zhengguo Wang et al. Distinctive momentum dependent charge-density-wave gap observed in CsV3Sb5 superconductor with topological Kagome lattice. *arXiv: 2104.05556* (2021).

[21] Zhonghao Liu et al. Temperature-induced band renormalization and Lifshitz transition in a kagome superconductor RbV3Sb5. *arXiv: 2104.01125* (2021).

[22] Yang Luo et al. Distinct band reconstructions in kagome superconductor CsV3Sb5. *arXiv: 2106.01248* (2021).

[23] Yong Hu et al. Rich Nature of Van Hove Singularities in Kagome Superconductor CsV3Sb5. *arXiv: 2106.05922* (2021).

[24] Yong Hu et al. Charge-order-assisted topological surface states and flat bands in the kagome superconductor CsV3Sb5. *arXiv: 2104.12725* (2021).

[25] Soohyun Cho et al. Emergence of new van Hove singularities in the charge density wave state of a topological kagome metal RbV3Sb5. *arXiv:2105.05117* (2021).

[26] Mingu Kang et al. Twofold van Hove singularity and origin of charge order in topological kagome superconductor CsV3Sb5. *arXiv: 2105.01689* (2021).

[27] Hailan Luo et al. Electronic Nature of Charge Density Wave and Electron-Phonon Coupling in Kagome Superconductor KV3Sb5. *arXiv: 2107.02688* (2021).

[28] Hengxin Tan et al. Charge density waves and electronic properties of superconducting kagome metals. *arXiv: 2103.06325* (2021).

[29] Jianzhou Zhao et al. Electronic correlations in the normal state of kagome superconductor KV3Sb5. *arXiv: 2103.15078* (2021).

[30] Chaoyu Chen, Shaolong He, Hongming Weng, Wentao Zhang, Lin Zhao, Haiyun Liu, Xiaowen Jia, Daixiang Mou, Shanyu Liu, Junfeng He, Yingying Peng, Ya Feng, Zhuojin Xie, Guodong Liu, Xiaoli Dong, Jun Zhang, Xiaoyang Wang, Qinjun Peng, Zhimin Wang, Shenjin Zhang, Feng Yang, Chuangtian Chen, Zuyan Xu, Xi Dai, Zhong Fang & X. J. Zhou. Robustness of topological order and formation of quantum well states in topological insulators exposed to ambient environment. *Proceedings of the National Academy of Sciences* **109**, 3694-3698, doi:10.1073/pnas.1115555109 (2012).

[31] M. S. Bahramy, P. D. C. King, A. de la Torre, J. Chang, M. Shi, L. Patthey, G. Balakrishnan, Ph Hofmann, R. Arita, N. Nagaosa & F. Baumberger. Emergent quantum confinement at topological insulator surfaces. *Nat Commun* **3**, 1159 (2012).

[32] Marco Bianchi, Richard C. Hatch, Jianli Mi, Bo Brummerstedt Iversen & Philip Hofmann. Simultaneous Quantization of Bulk Conduction and Valence States through Adsorption

[32] of Nonmagnetic Impurities on Bi_{2}Se_{3}. *Physical Review Letters* **107**, 086802 (2011).

[33] Marco Bianchi, Dandan Guan, Shining Bao, Jianli Mi, Bo Brummerstedt Iversen, Philip D. C. King & Philip Hofmann. Coexistence of the topological state and a two-dimensional electron gas on the surface of Bi2Se3. *Nat Commun* **1**, 128 (2010).

[34] Zhanyang Hao, et al. Multiple Symmetry-Protected Dirac Nodal Lines in A Quasi-One-Dimensional Semimetal. *arXiv preprint arXiv:2104.02221* (2021).

[35] Chaoyu Chen, José Avila, Shuopei Wang, Yao Wang, Marcin Mucha-Kruczyński, Cheng Shen, Rong Yang, Benjamin Nosarzewski, Thomas P. Devereaux, Guangyu Zhang & Maria Carmen Asensio. Emergence of Interfacial Polarons from Electron–Phonon Coupling in Graphene/h-BN van der Waals Heterostructures. *Nano letters* **18**, 1082-1087, doi:10.1021/acs.nanolett.7b04604 (2018).

[36] Chaoyu Chen, José Avila, Emmanouil Frantzeskakis, Anna Levy & Maria C Asensio. Observation of a two-dimensional liquid of Frohlich polarons at the bare SrTiO3 surface. *Nature Communications* **6**, 8585 (2015).

[37] J. Wang, J. Liu, Y. Xu, J. Wu, B.-L. Gu & W. Duan. Structural stability and topological surface states of the SnTe (111) surface. *Physical Review B* **89**, 125308, doi:10.1103/PhysRevB.89.125308 (2014).

[38] G. Kresse & J. Furthmüller. Efficient iterative schemes for ab initio total-energy calculations using a plane-wave basis set. *Physical Review B* **54**, 11169-11186, doi:10.1103/PhysRevB.54.11169 (1996).

[39] P. E. Blöchl. Projector augmented-wave method. *Physical Review B* **50**, 17953-17979, doi:10.1103/PhysRevB.50.17953 (1994).

[40] John P. Perdew, Kieron Burke & Matthias Ernzerhof. Generalized Gradient Approximation Made Simple. *Physical Review Letters* **77**, 3865-3868, doi:10.1103/PhysRevLett.77.3865 (1996).

[41] S. Grimme, J. Antony, S. Ehrlich & H. Krieg. A consistent and accurate ab initio parametrization of density functional dispersion correction (DFT-D) for the 94 elements H-Pu. *J Chem Phys* **132**, 154104, doi:10.1063/1.3382344 (2010).